\documentclass[slac_one]{revtex4}
\usepackage{graphicx}
\usepackage{fancyhdr}
\begin{document}

\title{Dynamical Symmetry Breaking and the Cosmological Constant Problem} 

\author{Philip D. Mannheim}
\affiliation{Department of Physics, University of Connecticut, Storrs, CT 06269, USA.~~philip.mannheim@uconn.edu}

\begin{abstract}
We outline a program with the potential to solve both the cosmological constant and quantum gravity problems within a single, comprehensive framework, one that is formulated entirely in four spacetime dimensions. The program is based on an interplay between the dynamical symmetry breaking of particle physics and a gravity theory, conformal gravity,  which possesses an underlying local conformal invariant structure. Central to the discussion is the recognition that when symmetry is broken dynamically in a conformal invariant theory, the cosmological constant is both induced by and constrained by zero-point fluctuations. We describe what  still needs to be done.
\end{abstract}

\maketitle

\thispagestyle{fancy}

\section{THE NEED FOR DYNAMICAL SYMMETRY BREAKING}

Ordinarily, one  thinks of the cosmological constant problem as a mismatch between particle physics wanting a large $\Lambda$ and standard cosmology needing a small one. However, this is not their only mismatch, with there being three other, equally significant, ones. Firstly, we note that in an $SU(3)\times SU(2)\times U(1)$ theory with massless fermions and gauge bosons the action is locally conformal invariant and the energy-momentum tensor $T^{\mu\nu}$ is traceless, save only for the wrong sign mass term in the Higgs potential. However, if particle symmetries are broken dynamically, the  Higgs field would only be a c-number order parameter in an effective Ginzburg-Landau Lagrangian which is output to dynamical symmetry breaking in a theory with no initial mass scales. Thus if weak interactions are broken in the same way as chiral symmetry is broken in strong interactions, we would be led to an elementary particle theory which would be  scale free at the level of the Lagrangian, and a $T^{\mu\nu}$ which would be traceless. Such a $T^{\mu\nu}$ could never be the source of standard Einstein gravity since it would force the Ricci scalar to be zero in every conceivable gravitational situation. In addition, the very use of the familiar double-well potential associated with a fundamental Higgs scalar field entails a second mismatch. Specifically, in flat space physics all that matters is the energy difference between the local maximum of this potential and the true degenerate minimum, and one can add on a constant to the double-well potential without modifying flat space physics at all. However, gravity does not couple to energy difference but to energy itself, and thus gravity is sensitive to where the zero of energy of the Higgs potential is, something that neither $SU(3)\times SU(2)\times U(1)$ nor standard gravity is currently able to specify. Thirdly, we note that in determining a particle physics expectation for $\Lambda$ in the first place, one ordinarily associates $\Lambda$ with the vacuum while associating the matter energy density $\rho_M$ with particle states with positive energy. Since these two quantities are to be associated with different classes of states, it is very difficult to relate or constrain them in any way.
 
As we shall show, all of the above concerns can be addressed if one dispenses with fundamental Higgs fields altogether, and instead takes all symmetries to be broken dynamically through fermion condensates in a theory with no fundamental mass scales whatsoever. In such a case the Higgs field would only emerge as c-number order parameter, and as such would not be produceable in a particle accelerator. With the source of gravity possessing no fundamental mass scales, the gravitational action itself would equally have to be locally conformal invariant, and thus uniquely be of the form $
I_W=-\alpha_g\int d^4x (-g)^{1/2}C_{\lambda\mu\nu\kappa} C^{\lambda\mu\nu\kappa}$ where $\alpha_g$ is a dimensionless coupling constant and $C^{\lambda\mu\nu\kappa}$ is the conformal Weyl tensor. The associated gravitational equations of motion for the metric are the fourth-order derivative (see e.g. \cite{Mannheim2006})
\begin{equation}
4\alpha_g W^{\mu\nu}=4\alpha_g [2C^{\mu\lambda\nu\kappa}_{\phantom{\lambda\mu\nu\kappa};\lambda;\kappa}
-C^{\mu\lambda\nu\kappa}R_{\lambda\kappa}]
=T^{\mu\nu},
\label{1}
\end{equation}
where both $W^{\mu\nu}$ and a conformal invariant $T^{\mu\nu}$ are kinematically traceless. 

As regards the problem of quantum gravity, we note that the problem is not that of an intrinsic mismatch between quantum mechanics and spacetime curvature per se. Rather, it is a problem for a particular theory of gravity, namely the standard Einstein theory. Specifically, because the standard Einstein-Hilbert action possesses a fundamental dimensionful scale, the Planck scale, quantum corrections to the theory are not renormalizable. Attempts to solve this problem have led to generalizing standard gravity to a ten-dimensional superstring theory, with there as yet being no evidence that any of the required six extra dimensions actually exist. However, the problem of making sense of a quantum theory with a dimensionful coupling constant is also met in the weak interactions, with it being the realization that the dimensionful $G_{\rm F}$ is not fundamental, and that the fundamental theory is instead based on massless fermions and gauge bosons which eventually led to the resolution of the problem. Thus to solve the problem of quantum gravitational corrections, we should try to make gravity look more like the other fundamental interactions, and thus also base gravity on a theory with no fundamental mass scales. Thus,  just as with the issue of the tracelessness of $T^{\mu\nu}$, we are again quite naturally led to consider conformal gravity, since with its dimensionless $\alpha_g$ as in (\ref{1}), the theory is renormalizable. Moreover, because of the required absence of fundamental mass scales, the theory is not allowed to possess any cosmological constant term at the level of the Lagrangian. 

Despite all this, conformal gravity has not been favored by the community, as it is thought to have a unitarity problem, with theories based on higher-derivative equations being thought to possess  ghost states.  However, a recent study \cite{Bender2007} of a fourth-order derivative quantum-mechanical system has shown that higher-derivative theories  have not been formulated properly. It turns out that the Hamiltonians of such theories are not Hermitian but are instead $PT$-symmetric. Consequently, it is incorrect to use the standard Dirac norm. Rather one must use an appropriately adapted  $PT$ norm, and when one does,  there are then no negative norm states at all and time evolution is unitary.
 
\section{IMPLICATIONS OF THE VANISHING OF THE TRACE OF ${\bf T^{\mu\nu}}$}
 
To identify what the implications of the vanishing of the trace of $T^{\mu\nu}$ might be, it is convenient to study the illustrative case of a massless four-fermion theory in dimension $D=2$  as that is the dimension in which the four-fermion theory is conformal invariant. In the flat spacetime case first, the Lagrangian is taken to be of the form $L =i\bar{\psi}\gamma^{\mu}\partial_{\mu}\psi -(g/2)[\bar{\psi}\psi]^2$, the equation of motion is given by $i\gamma^{\mu}\partial_{\mu}\psi -g[\bar{\psi}\psi]\psi=0$, and $T^{\mu\nu}$ is given by the traceless
\begin{equation}
T^{\mu\nu}=i\bar{\psi}\gamma^{\mu}\partial^{\nu}\psi -g^{\mu\nu}(g/2)[\bar{\psi}\psi]^2,\qquad 
T^{\mu}_{\phantom{\mu}\mu}=i\bar{\psi}\gamma^{\mu}\partial_{\mu}\psi -g[\bar{\psi}\psi]^2=0.
\label{2}
\end{equation}
In the mean-field approximation one looks for self-consistent, translation invariant, states $|S\rangle$ in which $\langle S|\bar{\psi}\psi | S\rangle =m/g$ and  $\langle S|[\bar{\psi}\psi -m/g]^2| S\rangle =0$, with the parameter $m$ being independent of the spacetime coordinates. In such states the $D=2$ fermion equation of motion and $T^{\mu\nu}$ are given as 
\begin{equation}
i\gamma^{\mu}\partial_{\mu}\psi -m\psi=0,\qquad \langle S|T^{\mu\nu}| S\rangle=\langle S|i\bar{\psi}\gamma^{\mu}\partial^{\nu}\psi| S\rangle -g^{\mu\nu}m^2/2g, \qquad \langle S|T^{\mu}_{\phantom{\mu}\mu}| S\rangle=m\langle S|\bar{\psi}\psi| S\rangle -m^2/g=0,
\label{3}
\end{equation}
with the mean-field approximation preserving tracelessness. If one incoherently sums $\langle S|i\bar{\psi}\gamma^{\mu}\partial^{\nu}\psi| S\rangle$ over modes which obey $\langle S|\bar{\psi}\psi | S\rangle =m/g$ (i.e. if one incoherently sums  over all momentum directions), $\langle S|i\bar{\psi}\gamma^{\mu}\partial^{\nu}\psi| S\rangle$ will take the form of a matter field perfect fluid $(\rho_m+p_m)U^{\mu}U^{\nu}+p_mg^{\mu\nu}$. On setting $m^2/2g=\Lambda$, we can thus set
\begin{equation}
\langle S|T^{\mu\nu}| S\rangle=(\rho_m+p_m)U^{\mu}U^{\nu}+p_mg^{\mu\nu} -g^{\mu\nu}\Lambda, \qquad \langle S|T^{\mu}_{\phantom{\mu}\mu}| S\rangle=p_m-\rho_m -2\Lambda=0.
\label{4}
\end{equation}
As we see, because the cosmological constant and the matter fluid are constructed from one and the same set of states, we are able to relate them, with the trace condition forcing $\Lambda$ to be neither bigger nor smaller than the matter fluid term, no matter how big $\Lambda$ itself might be. This relation stands in sharp contrast to the situation with a fundamental Higgs field, since there the scale associated with the double-well Higgs potential bears no relation whatsoever to the scale associated with the matter fluid. With regard to cosmological applications, we note that the above analysis holds independent of whether the state $|S\rangle$ is pure vacuum (i.e. just the filled negative energy Dirac sea) or includes positive energy states as well, i.e. states which are occupied in the cosmological case. The key to the $D=2$ tracelessness condition $p_m-\rho_m -2\Lambda=0$ is that all terms in it are evaluated in the same states.

As constructed, the above mean-field approximation takes care not only of the contribution of $\Lambda$ to the trace condition, but also of the zero-point energy contribution as well. As such, the zero-point energy problem arises because annihilation operators can precede creation operators in products of fermion fields, and thus yield a c-number anti-commutator term (i.e. $\langle \Omega |bb^{\dagger}|\Omega \rangle=\langle \Omega|(bb^{\dagger}+b^{\dagger}b)|\Omega \rangle$). Thus if we expand the massive fermion field as $\psi(x)=\sum _{\pm s}\int dp/(2\pi)^{1/2}(m/E_p)^{1/2}[b(p,s)u(p,s)e^{-ip\cdot x} +d^{\dagger}(p,s)v(p,s)e^{ip\cdot x}]$ as summed over spins and momenta, we find that the vacuum expectation value of the fermion kinetic energy term evaluates to $\langle \Omega|i\bar{\psi}\gamma^{\mu}\partial^{\nu}\psi |\Omega\rangle=-2\int dp p^{\mu}p^{\nu}/(2\pi E_p)$. Consequently, the energy density is given as $\langle \Omega|i\bar{\psi}\gamma^0\partial^0\psi |\Omega\rangle=-2\int dp E_p/2\pi$. We recognize the energy density as being none other than the familiar $\epsilon(m)=i\int d^2p/(2\pi)^2{\rm Tr}{\rm ln}[\gamma^{\mu}p_{\mu}-m+i\epsilon]$. In the ground state then the mean-field energy density is none other than the zero-point energy density. Now in flat space physics one ordinarily discards the zero-point energy by normal ordering. However, in the curved space case one cannot do so since gravity couples to energy itself and not to energy differences. For gravitational applications then one must keep the entire zero-point contribution. Nonetheless, in the trace we see that this term is then cancelled identically by the $m^2/g$ term, i.e. by none other than the induced cosmological constant term itself. In the mean-field approximation then, in the self-consistent vacuum the zero-point fluctuations serve to generate both a zero-point energy and a cosmological constant term which cancels it in the trace. That this has to happen is because the canonical fermion anti-commutator is conformal invariant, to thereby not affect the vanishing of the mean-field $T^{\mu}_{\phantom{\mu}\mu}$.

When the above analysis is made in a background gravitational field, the only change is that one introduces the fermion spin connection according to  $i\bar{\psi}\gamma^{\mu}(x)[\partial_{\mu}+\Gamma_{\mu}(x)]\psi$. If the background gravitational field is an expanding cosmology, the mean-field trace will still vanish but will now do so by having the $\langle S|i\bar{\psi}\gamma^{\mu}(x)[\partial^{\nu}+\Gamma^{\nu}(x)]\psi| S\rangle$ and $m^2/g=m \langle S|\bar{\psi}\psi| S\rangle$ terms redshift together. Since the $m^2/g$ term is dynamically induced, it is not a true constant in the sense that a fundamental $\Lambda$ would be, but instead adjusts to whichever states $|S\rangle$ are occupied in whatever cosmological background is chosen. A dynamically induced cosmological constant term thus adjusts with redshift.

In the dynamical symmetry breaking case there is a further set of states of interest, namely coherent states $|C\rangle$ in which the matrix element $\langle C|\bar{\psi}(x)\psi(x) |C\rangle$ is equal to $m(x)/g$ where $m(x)$ is now an extended, spacetime dependent configuration. Such states can be associated with the variational minimum of the vacuum functional $-\int d^Dx W[m(x)]=-\int d^Dx[-\epsilon[m(x)]+(1/2)Z[m(x)]\partial^{\mu}m(x)\partial_{\mu}m(x)+....]$ where $\epsilon[m(x)]$ is the vacuum energy as evaluated in the state with $m=m(x)$, with the variation in $m(x)$ causing $p_m-\rho_m$ to differ from $2\Lambda$ by the appropriate amount. The spacetime dependence of $m(x)$ will not be as big as the induced $\Lambda$ itself but only as big as the departure of $p_m-\rho_m -2\Lambda$ from zero. The variation in $m(x)$ can thus be small even if $\Lambda$ is huge. Moreover this will remain true even if the mean-field approximation is performed in curved space where $\partial^{\mu}m(x)\partial_{\mu}m(x)$ is schematically replaced by the conformally coupled  $\partial^{\mu}m(x)\partial_{\mu}m(x)-c m^2(x)R^{\alpha}_{\phantom {\alpha}\alpha}$ [$c$ is an appropriate constant which depends on the conformal weight of $m(x)$]. Thus in the trace, the Ricci scalar will couple not to $\Lambda$ itself but only to the departure of $-Z[m(x)][\nabla^{\mu}\nabla_{\mu}m(x)+cm(x)R^{\mu}_{\phantom{\mu}\mu}]+...\sim p_m-\rho_m-2\Lambda$ from zero. In the weak gravity case in which the Ricci scalar is small, the cancellation of $p_m-\rho_m$ with the $2\Lambda$ term will be almost as total as it is in the absence of gravity no matter how large the cosmological constant itself may be. In the dynamical symmetry breaking case then, the tracelessness of $T^{\mu\nu}$ forces gravity not to couple to $\Lambda$ itself but only to the difference $p_m-\rho_m-2\Lambda$, a quantity which is small when the matrix elements of  $i\bar{\psi}\gamma^{\mu}(x)[\partial_{\mu}+\Gamma_{\mu}(x)]\psi$ and $\bar{\psi}\psi$ are all evaluated in the same self-consistent states. In a situation such as this, there is no need to ever quench $\Lambda$ at all. It can be as large as particle physics has always expected it to be, and yet still lead to weak gravitational effects.

\section{IMPLICATIONS OF THE VANISHING OF ${\bf T^{\mu\nu}}$ ITSELF}

While the above discussion on the role of the trace condition is very general, in and of itself it does not completely resolve the cosmological constant problem since the trace only involves one particular combination of $\rho_m$, $p_m$ and $\Lambda$. Specifically, if we set $T^{\mu\nu}=(\rho_m+p_m)U^{\mu}U^{\nu}+p_mg^{\mu\nu}-\Lambda g^{\mu\nu}$ and impose a vanishing trace so that $3p_m-\rho_m-4\Lambda=0$ (in $D=4$ now), we can rewrite $T^{\mu\nu}$ as $T^{\mu\nu}=(\rho_m+p_m)(U^{\mu}U^{\nu} +g^{\mu\nu}/4)$, a quantity which is traceless for any value of $\rho_m+p_m$. Moreover, while the mean-field approximation leads to a vanishing trace, in and of itself it does not lead to a vanishing $\rho_m+p_m$, with there being no cancellation of the induced $\Lambda$ in $T^{\mu\nu}$ itself. To constrain $\rho_m+p_m$, we need to constrain $T^{\mu\nu}$ and not just its trace, and thus need to turn to a gravity theory which has the requisite conformal $T^{\mu\nu}$ as its source. Consequently, we are led to conformal gravity as based on (\ref{1}). 

Recalling that the $D=4$ Weyl tensor vanishes identically in a  Robertson-Walker (RW) geometry, from (\ref{1})  we see that in a conformal cosmological background the energy-momentum tensor  obeys $T^{\mu\nu}=0$ identically. Unlike standard cosmology then, conformal cosmology precisely knows where the zero of energy is. Moreover, even after early universe dynamical symmetry breaking phase transitions,  $T^{\mu\nu}$ will remain zero if the universe continues to be  RW. With the condition $T^{\mu\nu}=0$ holding both before and after early universe phase transitions, all of $\rho_m$, $p_m$ and $\Lambda$ are therefore constrained, with no single one of them being able to be overwhelmingly larger than any of the others.

Since the $T^{\mu\nu}=0$ condition is due to the gravitational equations of motion,  we cannot enforce it  without taking gravity itself into account. Thus to cancel the matter field zero-point fluctuations in $T^{\mu\nu}$ itself we will need to include the zero-point fluctuations of the gravitational field, and as long as there are no renormalization anomalies (see below), the overall cancellation of zero-point fluctuations (and thus of the induced $\Lambda$) in $T^{\mu\nu}$ is secured by the underlying conformal invariance of the theory. That there will be zero-point fluctuations in the gravitational field is due to the fact that even though the classical Weyl tensor vanishes in an RW geometry, it involves products of fields, to thus yield a zero-point contribution in the quantum theory. Moreover, the fermion spin connection involves products of vierbeins, and so it too can lead to zero-point gravitational fluctuations. The beauty of conformal gravity then is that because it is renormalizable it leads to controlled zero-point fluctuations in the gravitational field, fluctuations which are then constrained by the underlying conformal invariance so as to cancel the zero-point fluctuations in the matter fields. Moreover, since the graviton modes of the conformal theory are massless, their zero-point fluctuations do not couple to the trace of $T^{\mu\nu}$, to leave the matter mean-field zero-point cancellation in $T^{\mu}_{\phantom{\mu}\mu}$ intact as is.  (Since the $T^{\mu}_{\phantom{\mu}\mu}=0$ condition is imposed by variation with respect to the matter fields, gravity must respect it, with the graviton thus being obliged to be massless.)

\section{WHAT STILL NEEDS TO BE DONE}

To develop the program outlined here will require some quite technical additional steps. One needs to develop a full, non-perturbative treatment of radiative-correction-induced dynamical symmetry breaking in particle physics, one which needs to incorporate gravitational radiative corrections as well. One needs to address the fact that radiative corrections lead to conformal anomalies. One needs to extend the quantum-mechanical study of \cite{Bender2007} to the full quantum conformal gravity theory. Finally, one needs to ascertain what the predictions of the conformal theory for cosmology might then be. Nonetheless, demanding as these requirements might be, they are all well within conventional particle physics and involve no new particles (such as supersymmetric ones) or additional dimensions, and open up the possibility that one can resolve the cosmological constant and quantum gravity problems using what are quite conventional, albeit non-routine, ingredients.

As a first step toward doing all this, we recall \cite{Mannheim1974} a non-perturbative study of dynamical symmetry breaking in QED with zero bare mass fermions. In \cite{Mannheim1974} it was assumed that $\alpha$ was at a zero of the renormalization group $\beta$ function, since at such a zero the conformal invariance that is broken by radiative corrections gets restored with anomalous dimensions. (One presumably would have to require that conformal gravity be at a zero of its $\beta$ function too, since it is otherwise very difficult to attach any significance to conformal invariance in the presence of radiative corrections.) In the study of \cite{Mannheim1974} it was found that if the dimension of the fermion composite bilinear $\bar{\psi}\psi$ is reduced from three to two, the vacuum would then undergo dynamical symmetry breaking and generate a fermion mass. The need for such a deep reduction in dimension is because the very act of softening the short distance behavior (dimension less than canonical) causes the theory  to be more divergent in the infrared, with this very deep reduction then being found to generate long range order and a dynamical double-well potential analogous to the one found in mean-field theory.  
The author wishes to thank Dr. J. Polchinski and Dr. E. Witten for helpful comments.


\begin{thebibliography}{9}  

\bibitem{Mannheim2006} P. D. Mannheim, Prog. Part. Nucl. Phys. 56, 340 (2006). 

\bibitem{Bender2007} C. M. Bender and P. D. Mannheim, Phys. Rev. Lett. 100, 110402 (2008); 
Phys. Rev. D 78, 025022 (2008). 

\bibitem{Mannheim1974} P. D. Mannheim, Phys. Rev. D 10, 3311 (1974); Phys. Rev. D 12, 1772 (1975); Nucl. Phys. B 143, 285 (1978).



\end{thebibliography}
\end{document}